\documentclass[12pt]{iopart}

\usepackage{amsgen}
\usepackage{amssymb}
\usepackage{epsfig}
\usepackage{setstack}

\begin{document}

\title{Further Exact Cosmological Solutions to Higher-Order Gravity Theories.}

\author{T Clifton and John D Barrow}

\address{DAMTP, Centre for Mathematical Sciences, University of
  Cambridge,  Wilberforce Road, Cambridge, CB3 0WA, UK.}

\eads{\mailto{\mailto{T.Clifton@datmp.cam.ac.uk},
\mailto{J.D.Barrow@damtp.cam.ac.uk}}}


\begin{abstract}

We investigate the effect of deviations from general relativity on approach
to the initial singularity by finding exact cosmological solutions to a wide
class of fourth-order gravity theories. We present new anisotropic vacuum
solutions of modified Kasner type and demonstrate the extent to which they
are valid in the presence of non-comoving perfect-fluid matter fields. The
infinite series of Mixmaster oscillations seen in general relativity will
not occur in these solutions, except in unphysical cases.

\end{abstract}

\maketitle

\section{Introduction}

There have been numerous studies in the literature on generalisations of the
usual Einstein-Hilbert action of general relativity to more complicated
functions of the curvature (see refs. in \cite{Bar83,schmidt}). The
motivation for such studies comes from a variety of different sources
ranging from attempts to include quantum effects in the gravitational action 
\cite{Stelle} to the investigation of phenomena that are presently
inadequately explained in the standard model, such as dark energy \cite%
{dark1,dark2}, and violations of the cosmic no hair theorem \cite{bher,bar}. Of
particular interest is the behaviour of these modified theories of gravity
in the high-curvature limit, where quantum corrections to general relativity
are expected to become important and their influence on the occurrence of
singularities and possible bounces of the universe at high curvatures can be
studied \cite{ruz}. It is the investigation of these modified theories of
gravity on approach to the initial cosmological singularity that concerns us
here. Previous studies have focussed on the behaviour of isotropic
cosmologies but, as we know from the situation in general relativity, their
behaviour can be misleading. Anisotropies diverge faster than isotropic
densities at high curvatures and will generally dominate the behaviour of
the cosmology at early (and in some cases even late) times. The most
important anisotropic solutions of general relativistic cosmologies are the
vacuum Kasner solutions \cite{kas} and their fluid-filled counterparts that
form Type I of the Bianchi classification of three-dimensional homogeneous
spaces. These universes are geometrically special, in vacuum (or
perfect-fluid) cases they are defined by just one (or two) free constant(s)
compared to the four (eight) that specify the most general spatially
homogeneous vacuum (or perfect fluid) solutions. However, they have proved
to provide an excellent dynamical description of the evolution of the most
general models over finite time intervals. The chaotic vacuum Mixmaster
universe of Bianchi Type IX undergoes a infinite sequence of chaotic
space-time oscillations on approach to its initial or final singularities
which is well approximated by a sequence of different Kasner epochs which
form a Poincar\'{e} return mapping for the chaotic dynamical system \cite%
{bkl, misner, jb1, jb2, chern, rend}.

The way in which the Kasner metric has played a central role in the
elucidation of the existence and structure of anisotropic cosmological
models and their singularities in general relativity makes it the obvious
starting point for an extension of that understanding to cosmological
solutions of higher-order gravity theories. In an earlier Letter \cite{Bar06}
we have reported the discovery of a new class of exact Kasner-like solutions
for gravity theories derived from Lagrangians that are an arbitrary power of
the scalar curvature. In this paper we will generalise that study to a much
wider class of higher-order Lagrangian theories. We will determine the
conditions for the existence of Kasner solutions and find their exact forms.
In some cases these solutions are required to be isotropic and correspond to
exact Friedmann-Robertson-Walker (FRW) vacuum solutions with zero spatial
curvature. By studying gravity theories whose Lagrangians are derived from
arbitrary powers of the curvature invariants we are able to find simple
exact solutions. Past studies have usually focussed on the addition of
higher-order curvature terms to the Einstein-Hilbert Lagrangian. This
results in enormous algebraic complexity and exact solutions cannot be
found. The results presented here provide a tractable route into
understanding the behaviours of anisotropic cosmological models in
situations where the higher-order curvature corrections are expected to
dominate. The solutions we present are vacuum solutions but we provide a
simple analysis which determines when the introduction of perfect fluids
with non-coming velocities has a negligible effect on the cosmological
evolution at early times.

\section{Field Equations}

We consider field equations derived from an arbitrary analytic function of
the three possible linear and quadratic contractions of the Riemann
curvature tensor; $R$, $R_{ab}R^{ab}$ and $R_{abcd}R^{abcd}$ \cite{dewitt}.
The weight-zero Lagrangian density for such theories is 
\begin{equation}
\mathcal{L}_{G}=\chi^{-1}\sqrt{g}f(X,Y,Z)  \label{density}
\end{equation}%
where $\chi $ is a constant and $f(X,Y,Z)$ is an arbitrary function of the
variables $X$, $Y$ and $Z$ defined as $X=R$, $Y=R_{ab}R^{ab}$ and $%
Z=R_{abcd}R^{abcd}$. We ignore the boundary terms, for a discussion of which
see ref. \cite{BM}. The field equations derived from the variation of the
corresponding action are \cite{BM, Cli05} 
\begin{eqnarray}
P^{ab} &=&-\frac{1}{2}fg^{ab}+f_{X}R^{ab}+2f_{Y}R^{ac}R_{%
\;c}^{b}+2f_{Z}R^{acde}R_{\;cde}^{b}  \label{fequations} \\
&&\qquad -\square
(f_{Y}R^{ab})-g^{ab}(f_{Y}R^{cd})_{;cd}+2(f_{Y}R^{c(a})_{;\;\;c}^{\;b)} 
\nonumber \\
&&\qquad -f_{X;cd}(g^{ab}g^{cd}-g^{ac}g^{bd})-4(f_{Z}R^{d(ab)c})_{;cd} 
\nonumber \\
&=&\frac{\chi }{2}T_{ab}  \nonumber
\end{eqnarray}%
where a subscript $X$, $Y$ or $Z$ denotes differentiation with respect to
that quantity.

We will be looking for spatially homogeneous, vacuum solutions of Bianchi
type I, described by the line--element 
\begin{equation}
ds^{2}=-dt^{2}+\sum_{i=1}^{3}t^{2p_{i}}dx_{i}^{2}  \label{kasner}
\end{equation}%
where $p_{1}$, $p_{2}$ and $p_{3}$ are constants to be determined. For the
special case $p_{1}=p_{2}=p_{3}$ these solutions correspond to spatially
flat FRW metrics, which will be found to exist for various higher-order
theories both in vacuum (where none exist in general relativity) and in the
presence of a perfect fluid.

We can determine the number of independently arbitrary functions of three
space variables that will characterise the \textit{general} vacuum solution
of the field equations on a Cauchy surface of constant time in these
higher-order theories. The field equations are in general $4^{th}$ order in
time; so if we choose a synchronous reference system then we need 6
functions each for the symmetric $3\times 3$ tensors $g_{\alpha \beta }$, $%
\dot{g}_{\alpha \beta }$, $\dddot{g}_{\alpha \beta }$ and $\ddddot{g}%
_{\alpha \beta }$. This gives 24 functions, but they may be reduced to 20 by
using the 4 constraint equations, and finally again to 16 by using the 4
coordinate covariances. If a perfect fluid were included as a matter source
the final number would rise to 20 due to the inclusion of the density and 3
non-comoving velocity components in the initial data count. We note that in
general relativity the general vacuum solution is prescribed by 3 arbitrary
functions of 3 space variables. In general relativity, the Kasner vacuum
solution is prescribed by one free constant, as the three $p_{i}$ satisfy
two algebraic constraints.

\section{Exact Solutions}

For a realistic theory we should expect the dominant term of the analytic
function $f(X,Y,Z)$ to be of the Einstein-Hilbert form in the Newtonian
limit. However, there is no reason to expect such a term to dominate in the
high curvature limit - in fact, this is the limit in which quantum
corrections should become dominant. We therefore allow the dominant term in
a power series expansion of $f(X,Y,Z)$ to be of the form $R^{n}$, $%
(R_{ab}R^{ab})^{n}$ or $(R_{abcd}R^{abcd})^{n}$ on approach to the
singularity, where $n$ is a constant. These three different cases will be
investigated separately below.

\subsection{$f=R^n$}

In a previous work \cite{Bar06} we found Bianchi type I solutions to the
theory defined by the choice $f=f(X)=R^{n}$. These solutions have a
particularly simple form and they could be used to show that anisotropic
universes in such theories do not exhibit an infinite sequence of Mixmaster
oscillations on approach to the initial singularity, if $n>1$. We also
include these solutions here for completeness.

Substituting $f=f(X)=R^{n}$ into the field equations (\ref{fequations})
together with the line-element (\ref{kasner}) gives the two independent
algebraic constraints, 
\[
(2n^{2}-4n+3)P+(n-2)Q-3(n-1)(2n-1)=0 
\]
and 
\[
2(n^{2}-1)P+(n-2)(P^{2}-Q)-3(n-1)^{2}(2n-1)=0, 
\]
where we have defined 
\begin{equation}  \label{Xn1}
P \equiv \sum_{i} p_{i} \quad and \quad Q \equiv \sum_{i}
p_{i}^{2}.
\end{equation}

The constraint equations have two classes of solution. The first is given by 
\begin{eqnarray*}
&&P=\frac{3(n-1)(2n-1)}{(2-n)} \\
&&Q=\frac{P^{2}}{3}
\end{eqnarray*}%
which is only solved by the isotropic solution $p_{1}=p_{2}=p_{3}=P/3$. This
is the zero-curvature isotropic vacuum universe found by Bleyer and Schmidt 
\cite{schmidt, schmidt1}. It can be seen that these isotropic cosmologies
are valid for all $n\neq 2$ and correspond to expanding universes for $n<1/2$
or $1<n<2$ and to collapsing universes for $1/2<n<1$ or $2<n$.

\begin{figure}[tbp]
\label{Xn} \centerline{\epsfig{file=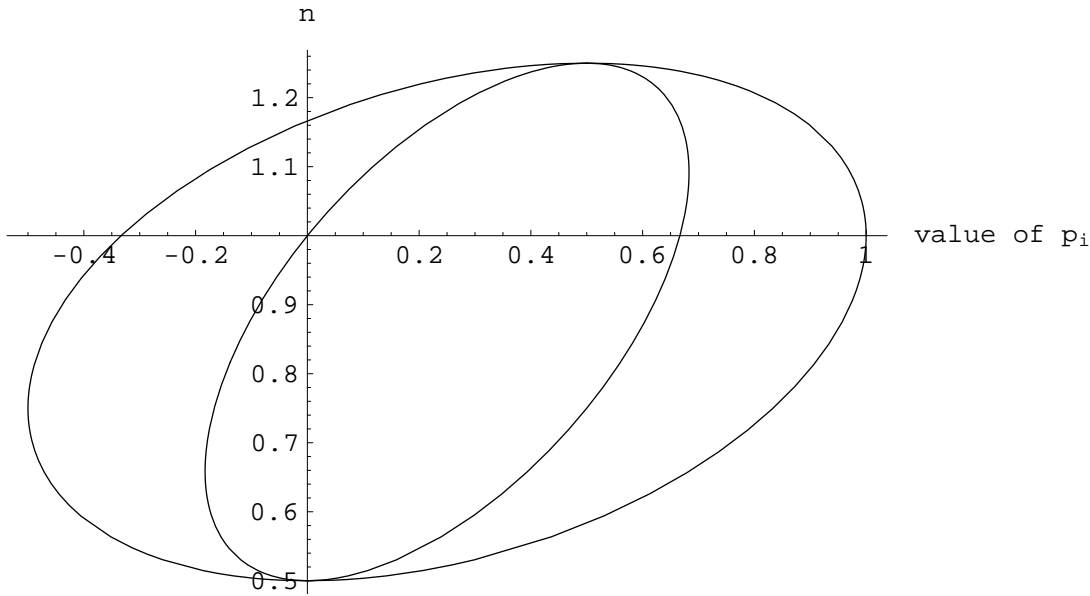,height=9cm}}
\caption{\textit{For the solution (\protect\ref{Xnkasner}), the intervals in
which the Kasner indices $p_{i}$ are allowed to lie can be read off from
this graph. For any value of $n$ in the range $\frac{1}{2}\leqslant
n\leqslant \frac{5}{4}$ a horizontal line is drawn; the boundaries of the
intervals in which the $p_{i}$ lie are then given by the four points at
which the horizontal line crosses the two closed curves. For $n=1$ these
boundaries can be seen to be $-\frac{1}{3}$, $0$, $\frac{2}{3}$ and $1$, as
expected for the Kasner solution of general relativity.}}
\end{figure}

The second class of solutions to (\ref{Xn1}) is given by 
\begin{eqnarray}
&&P=2n-1  \label{Xnkasner} \\
&&Q=(2n-3)(1-2n).  \nonumber
\end{eqnarray}%
This class of solutions represents a generalisation of the Kasner metric of
general relativity and was the basis of the study \cite{Bar06}. It was shown
in \cite{Bar06} that this class of solutions only exists for $n$ in the
range $1/2\leqslant n\leqslant 5/4$ and that the values of the constants $%
p_{i}$ must then lie within the ranges 
\begin{eqnarray*}
2n-1-2\sqrt{(2n-1)(5-4n)} &\leqslant &3p_{1}\leqslant 2n-1-\sqrt{(2n-1)(5-4n)%
}, \\
2n-1-\sqrt{(2n-1)(5-4n)} &\leqslant &3p_{2}\leqslant 2n-1+\sqrt{(2n-1)(5-4n)}%
, \\
2n-1+\sqrt{(2n-1)(5-4n)} &\leqslant &3p_{3}\leqslant 2n-1+2\sqrt{(2n-1)(5-4n)%
},
\end{eqnarray*}%
where it has been assumed without loss of generality that $p_{1}\leqslant
p_{2}\leqslant p_{3}$. These ranges are shown on Figure \ref{Xn} and can be
read off by drawing a horizontal line of constant $n$ on the plot. The four
points at which the horizontal line crosses the two closed curves gives the
four boundary values for the allowed ranges of the constants $p_{i}$. It can
be seen that the points at which the curves cross the abscissa, which
corresponds to $n=1$, give the boundary values $-1/3\leqslant p_{1}\leqslant
0\leqslant p_{2}\leqslant 2/3\leqslant p_{3}\leqslant 1$, in agreement with
the Kasner solution of general relativity. For $n>1/2$ these solutions
correspond to expanding universes with a curvature singularity at $t=0$. For 
$n=1/2$ the only solution is Minkowski spacetime. 

For a universe filled with a perfect fluid having an equation of state $%
p=(\gamma -1)\rho $, $\gamma $ constant, relating the fluid pressure $p$ to
its density $\rho ,$ the field equations (\ref{fequations}) have the
isotropic solution 
\begin{equation}
p_{1}=p_{2}=p_{3}=\frac{2n}{3\gamma },
\end{equation}%
for $\gamma \neq 0$. This reduces to the spatially-flat FRW solution of
general relativity in the limit $n\rightarrow 1$. For $n>0,$ these isotropic
cosmologies are expanding, and for $n<0$ they are contracting, with $n=0$
giving Minkowski spacetime. The stability and observational consequences of
cosmologies of this type were investigated in \cite{Cli05, Car04}, where
primordial nucleosynthesis and the microwave background were used to impose
observational constraints on the admissible values of $n$.

\subsection{$f=(R_{ab} R^{a b})^n$}

Substituting $f=f(Y)=(R_{ab}R^{ab})^{n}$ into the field equation (\ref%
{fequations}) along with the metric ansatz (\ref{kasner}) gives the two
independent equations, 
\begin{eqnarray}
Y^{n-1}\Bigl( &&(P^{2}+Q-4PQ+P^{2}Q+Q^{2})-  \label{Yn1} \\
&&\;2(3P^{2}+P^{3}+3Q-9PQ+2Q^{2})n+2(4P^{2}+4Q-8PQ)\Bigr)=0  \nonumber
\end{eqnarray}%
and 
\begin{eqnarray}
Y^{n-1}\Bigl( &&(24P-2P^{2}-2P^{3}-30Q+10PQ)n+(8P^{2}-32P+24Q)n^{2}
\label{Yn2} \\
&&\;\;\;+(P^{2}-4P+2P^{3}+9Q-10PQ-P^{2}Q+3Q^{2})\Bigr)=0,  \nonumber
\end{eqnarray}%
where $P$ and $Q$ are defined as before. Equations (\ref{Yn1}) and (\ref{Yn2}%
) have five classes of solutions.

The first of these classes is given by 
\[
P=Q=0.
\]%
This class of solutions is only satisfied by $p_{1}=p_{2}=p_{3}=0$, which is
Minkowski spacetime. It can be seen from (\ref{Yn1}) and (\ref{Yn2}) that as
this solution corresponds to $Y=0$ it only exists for $n\geqslant 0$ (for $%
n<0$ the premultiplicative factor in (\ref{Yn1}) and (\ref{Yn2}) causes the
left-hand side of those equations to diverge).

The second class of solutions is given by 
\[
P=Q=1.
\]%
This is just the Kasner solution of general relativity, for which the values
of the constants $p_{i}$ are constrained to lie within the ranges $%
-1/3\leqslant p_{1}\leqslant 0\leqslant p_{2}\leqslant 2/3\leqslant
p_{3}\leqslant 1$. Again, this solution corresponds to $Y=0$ and so is only
valid for $n\geqslant 0$.

The third class of solutions is given by 
\begin{eqnarray*}
&&P=\frac{3(1-3n+4n^{2})\pm \sqrt{3(-1+10n-5n^{2}-40n^{3}+48n^{4})}}{2(1-n)}
\\
&&Q=\frac{P^{2}}{3}.
\end{eqnarray*}%
The only solution belonging to this class corresponds to an isotropic and
spatially flat vacuum FRW cosmology. The values of the constants $p_{i}$ are
all equal to $P/3$ in this case and the solution is valid for all $n\neq 1$.

\begin{figure}[tbp]
\label{Yn} \centerline{\epsfig{file=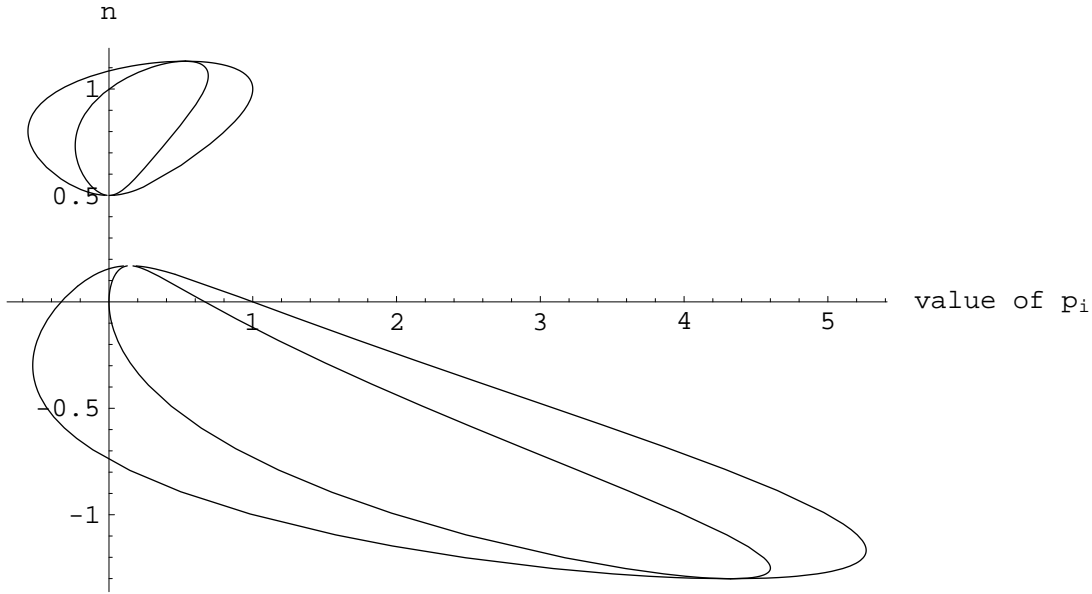,height=9cm}}
\caption{\textit{For the solution (\protect\ref{Ynkasner}) the intervals in
which the Kasner indices $p_{i}$ are constrained to lie can be read off this
graph, as with figure \protect\ref{Xn}. A horizontal line of constant $n$ is
drawn; the boundaries of the intervals in which the $p_{i}$ lie are then
given by the four points at which the horizontal line crosses the curves.
For $n=1$ these boundaries can be seen to be $-\frac{1}{3}$, $0$, $\frac{2}{3%
}$ and $1$, as expected for the Kasner solution of general relativity.}}
\end{figure}

The fourth class of solutions is given by 
\begin{eqnarray}
&&P=(1-2n)^{2}  \label{Ynkasner} \\
&&Q=1-8n+16n^{2}-8n^{3}.  \nonumber
\end{eqnarray}%
The solutions belonging to this class are anisotropic cosmologies which
reduce to the standard Kasner form, $P=Q=1$, in the limit $n\rightarrow 1$.
For this class of solutions, the values of the constants $p_{i}$ are in
general constrained to lie within the ranges 
\begin{eqnarray*}
(1-2n)^{2}-2A &\leqslant &3p_{1}\leqslant (1-2n)^{2}-A \\
(1-2n)^{2}-A &\leqslant &3p_{2}\leqslant (1-2n)^{2}+A \\
(1-2n)^{2}+A &\leqslant &3p_{3}\leqslant (1-2n)^{2}+2A
\end{eqnarray*}%
where $A=\sqrt{(1-2n)(1-6n+4n^{3})}$ and the $p_{i}$ have again been ordered
so that $p_{1}\leqslant p_{2}\leqslant p_{3}$. These boundaries are shown in
figure \ref{Yn} which can be read in the same way as figure \ref{Xn}, by
taking a horizontal line of constant $n$ and noting the four points at which
that line intersects the curves. These intercepts determine the allowed
intervals for the values of the $p_{i}$. 
For real-valued $p_{i}$, the value of $n$ must lie either in the range $%
n_{1}\leqslant n\leqslant n_{2}$ or in the range $1/2\leqslant n\leqslant
n_{3}$, where $n_{1}$, $n_{2}$ and $n_{3}$ are the roots of the cubic
polynomial $1-6n+4n^{3}=0$ and are chosen such that $n_{1}<n_{2}<n_{3}$.
These generalisations of the Kasner metric always correspond to expanding
cosmologies, independent of the value of $n$.

The fifth class of solutions to (\ref{Yn1}) and (\ref{Yn2}) is given by 
\begin{eqnarray*}
&&P=4n-1 \\
&&Q=-3+12n-8n^{2}\pm 2(1-2n)\sqrt{2(1-4n+2n^{2})}.
\end{eqnarray*}%
This class describes complex-valued $p_{i}$ for all values of $n$ (except
for $n=1/4$, for which this class of solutions reduces to the first class)
and is therefore of limited interest.

The isotropic solution for a universe filled with a fluid with equation of
state $p=(\gamma -1)\rho $ is given by the choice 
\[
p_{1}=p_{2}=p_{3}=\frac{4n}{3\gamma }.
\]%
This solution reduces to the spatially-flat FRW cosmology of general
relativity in the limit $n\rightarrow 1/2$ and corresponds to an expanding
universe for $n>0$ and to a contracting universe for $n<0$.

\subsection{$f=(R_{a b c d} R^{a b c d})^n$}

Substituting $f=f(Z)=(R_{abcd}R^{abcd})^{n}$ and (\ref{kasner}) into the
field equations (\ref{fequations}) gives the two independent equations 
\begin{eqnarray}
Z^{n-1}\Biggl( &&\Bigl(\frac{P^{2}Q}{2}-\frac{P^{4}}{12}-Q-\frac{3Q^{2}}{4}%
+2S-\frac{2PS}{3}\Bigr)  \label{Zn1} \\
&&+\Bigl(\frac{P^{4}}{3}+6Q+2PQ-2P^{2}Q+3Q^{2}-10S+\frac{2PS}{3}\Bigr)n 
\nonumber \\
&&\;+8(S-Q)n^{2}\Biggr)=0  \nonumber
\end{eqnarray}%
and 
\begin{eqnarray}
Z^{n-1}\Biggl( &&\Bigl(\frac{P^{4}}{4}-P-\frac{3P^{2}}{2}-\frac{P^{3}}{2}+%
\frac{13Q}{2}+\frac{7PQ}{2}-\frac{3P^{2}Q}{2}+\frac{9Q^{2}}{4}-8S\Bigr)
\label{Zn2} \\
&&\;+(6P+4P^{2}-16Q-2PQ+8S)n  \nonumber \\
&&\;+8(Q-P)n^{2}\Biggr)=0  \nonumber
\end{eqnarray}%
where $P$ and $Q$ are defined as before and we have also now defined

\[
S \equiv \sum_{i}p_{i}^{3}. 
\]
Equations (\ref{Zn1}) and (\ref{Zn2}) have four different classes of
solution.

The first class of solutions is given by 
\[
P=Q=S=0. 
\]%
The only solution that belongs to this class is Minkowski spacetime. As $%
R_{abcd}R^{abcd}=0$ for Minkowski spacetime it can only be a solution for $%
n\geqslant 0$, due to the premultiplier in equations (\ref{Zn1}) and (\ref%
{Zn2}).

The second class of solutions is given by 
\[
P=Q=S=1. 
\]%
The only solution that belongs to this class corresponds to $p_{1}=p_{2}=0$
and $p_{3}=1$. Making the coordinate transformations $\bar{z}=t\sinh z$ and $%
\bar{t}=t\cosh z$ \cite{landau} allows the line-element (\ref{kasner}) to
then be written in the form 
\[
ds^{2}=-d\bar{t}^{2}+dx^{2}+dy^{2}+d\bar{z}^{2} 
\]%
which is clearly Minkowski spacetime again. This is only a solution for $%
n\geq 0$, as with the first class of solutions.

The third class of solutions to (\ref{Zn1}) and (\ref{Zn2}) is given by 
\begin{eqnarray*}
&&P=\frac{3(1-2n+4n^{2}\pm \sqrt{-1+10n-16n^{2}+16n^{4}}}{2(1-n)}, \\
&&Q=\frac{P^{2}}{3}, \\
&&S=\frac{P^{3}}{9},
\end{eqnarray*}%
which only has the isotropic and spatially flat vacuum FRW cosmology as a
solution, where $p_{1}=p_{2}=p_{3}=P/3$. This is a solution for all $n\neq 1$%
.

The fourth and last class of solutions is given by 
\begin{eqnarray*}
&&P=4n-1 \\
&&Q=\frac{1}{3}\{16n^{2}-8n-1\pm 4\sqrt{2[n(1-2n)+S(1-n)]}\}.
\end{eqnarray*}%
This class of solutions corresponds to anisotropic metrics with $Z=0$ and is
unusual in that it cannot be expressed in the form $P=$ constant and $Q=$
constant. This feature means that the standard picture of a plane
intersecting an ellipsoid is no longer a valid one for this class. Solutions
in this class do not appear to have any range of $n$ for which the constants 
$p_{i}$ take real values, and so are of limited physical interest.

The isotropic solution for a universe filled with a fluid with equation of
state $p=(\gamma -1)\rho $ is given by 
\[
p_{1}=p_{2}=p_{3}=\frac{4n}{3\gamma }.
\]%
As before, this reduces to the spatially flat FRW solution of general
relativity in the limit $n\rightarrow 1/2$ and corresponds to an expanding
universe for $n>0$ and to a contracting universe for $n<0$.

\section{Investigation of the effects of matter}

We have identified above two generalisations of the Kasner metric of general
relativity which are solutions to the scale-invariant theories of gravity we
are investigating. The first of these is a solution to the theory defined by 
$\mathcal{L}=R^{n}$ and was the subject of investigation in \cite{Bar06};
the second is a new solution to the theory $\mathcal{L}=(R_{ab}R^{ab})^{n}$.
These are the first exact anisotropic solutions to be found for higher-order
gravity theories. In this section we will investigate some of the properties
of these cosmologies.

The behaviour of these solutions is of particular interest when considering
the Bianchi type VIII or type IX `Mixmaster' cosmologies. The field
equations for these cosmologies can be cast into the form of the equations
of motion of a particle moving inside an exponentially steep triangular
potential well with open channels in the corners \cite{misner}. The three
steep-sided walls are created by the 3-curvature anisotropies. In the region
where the potential is negligible (far from the walls) the behaviour of the
solutions approaches that of the Kasner metric. As the exponentially steep
potential wall is approached the universe `particle' is reflected and
re-enters a Kasner--like regime with the Kasner indices $p_{i}$
systematically changed to some new values by the rule of reflection from the
potential wall. In general relativistic cosmologies of Bianchi types VIII
and IX this process is repeated an infinite number of times as the
singularity is approached \cite{bkl, misner, jb1, jb2, chern, rend} so long
as matter obeys an equation of state with $p<\rho $. After reflection from
the potential wall the Kasner index that was previously negative is permuted
to a positive value and the lowest-valued positive Kasner index is permuted
to a negative value. This is repeated \textit{ad infinitum} in the general
relativistic cosmology as one of the Kasner indices must be kept negative
while the other two are positive. In the generalisations of the Kasner
metric presented above this is no longer the case; for some solutions it is
possible for all of the Kasner indices to be positive. If the Kasner indices
are being permuted in a chaotic fashion for long enough then, eventually,
they will end up in such a configuration. Once this occurs the oscillations
will end, as all spatial dimensions will be contracting, and the singularity
will reached monotonically without further oscillation of the scale factors 
\cite{BK, Bar78}.

From figure \ref{Xn} it can be seen that for solution (\ref{Xnkasner}) all
Kasner indices can be made positive for values of $1<n<5/4$, in the $%
\mathcal{L}=R^{n}$ theories. Similarly, from figure \ref{Yn} it can be seen
that for the solution (\ref{Ynkasner}), with $n_{1}<n<n_{2}$ or $1<n<n_{3},$
that all the Kasner indices can be made positive, in the theory $\mathcal{L}%
=(R_{ab}R^{ab})^{n}$ (where the $n_{i}$ are defined as before). For both of
these solutions, with these ranges of $n$, there is therefore no infinite
series of chaotic Mixmaster oscillations as the singularity is approached.
However, the solutions to both of these theories are still expected to
exhibit an infinite number of chaotic oscillations if $1/2<n<1$, as at least
one of the Kasner indices must be kept negative whilst another is kept
positive. A separate detailed analysis is require to determine where there
are differences between this behaviour and the chaotic oscillations found in
general relativity.

It now remains to show that the vacuum solutions found above are the
asymptotic attractor solutions in the presence of non-comoving matter
motions as $t\rightarrow 0$. This analysis will follow closely that of \cite%
{Bar06} and \cite{landau}. We aim to show that the fluid stresses diverge
more slowly than the vacuum terms as $t\rightarrow 0$ and so produce
negligible metric perturbations to an anisotropic Kasner universe. If this
is the case, then the vacuum solutions above can indeed be considered as the
asymptotic attractor solutions even in the presence of matter. Matter will
just be carried along by the expansion and behave like a test fluid.

We now consider a perfect fluid with equation of state $p=(\gamma -1)\rho $
where $1\leqslant \gamma <2$ and non-zero 4-velocity components, $u^{i}$
normalised so that $u^{i}u_{i}=1$. The conservation equations ${T^{ab}}%
_{;b}=0$ on the metric background (\ref{kasner}) can then be written in the
form \cite{landau,Bar78,Bar77} 
\begin{eqnarray*}
&&\frac{\partial }{\partial x^{i}}(t^{p}\rho ^{\frac{1}{\gamma }}u^{i})=0 \\
&&\gamma \rho u^{k}(u_{i,k}-\frac{1}{2}u^{l}g_{kl,i})=-\frac{1}{3}\rho
_{,i}-(\gamma -1)u_{i}u^{k}\rho _{,k}.
\end{eqnarray*}%
Neglecting spatial derivatives with respect to time derivatives these
equations integrate to 
\begin{eqnarray*}
t^{p}u_{0}\rho ^{\frac{1}{\gamma }} &=& const. \\
u_{\alpha }\rho ^{\frac{(\gamma -1)}{\gamma }} &=& const.
\end{eqnarray*}%
The neglecting of spatial derivatives means that these equations are valid
on scales larger than the particle horizon as in the velocity-dominated
approximation (although we have not so far restricted the fluid motions to
be non-relativistic) \cite{landau, els, tomita}.

From the second of the integrals above, it can be seen that all the
covariant components of the spatial 3-velocity, $u_{\alpha }$, are
approximately equal. This is not true of the contravariant components as the
Kasner indices, $p_{i}$, in the metric elements used to raise indices are
not equal in these solutions. The contravariant component that diverges the
fastest, and dominates the others in the $t\rightarrow 0$ limit, is
therefore $u^{3}=u_{3}t^{2p_{3}}$, as $p_{3}$ is the largest of the $p_{i}$.
If the 4-velocity is normalised, so that $u_{a}u^{a}=1$, and the
contravariant 3-velocity component $u^{3}$ diverges the fastest as $%
t\rightarrow 0,$ then we must have $u_{0}u^{0}\sim
u_{3}u^{3}=(u_{3})^{2}t^{-2p_{3}}$ in that limit. The integrated
conservation equations above can now be solved approximately in this limit
to give 
\begin{eqnarray*}
&&\rho \sim t^{-\gamma (p_{1}+p_{2})/(2-\gamma )} \\
&&u_{\alpha }\sim t^{(p_{1}+p_{2})(\gamma -1)/(2-\gamma )}
\end{eqnarray*}%
as $t\rightarrow 0$.

It is now possible to calculate the leading-order contributions to the
energy-momentum tensor, $T_{b}^{a}=(\rho +p)u^{a}u_{b}-p\delta _{b}^{a}$, as
being 
\begin{eqnarray*}
&T_{0}^{0}\sim \rho u^{0}u_{0}&\sim t^{-P-p_{3}} \\
&T_{1}^{1}\sim \rho &\sim t^{-\gamma (P-p_{3})/(2-\gamma )} \\
&T_{2}^{2}\sim \rho u^{2}u_{2}&\sim t^{-2p_{2}-P+p_{3}} \\
&T_{3}^{3}\sim \rho u^{3}u_{3}&\sim t^{-P-p_{3}}.
\end{eqnarray*}%
The component which diverges the fastest here as $t\rightarrow 0$ is $%
T_{3}^{3}\sim t^{-P-p_{3}}$, for general $\gamma $. For the case $\gamma =%
\frac{4}{3}$ where the 4-velocity of the fluid is comoving, $u_{i}=\delta
_{i}^{0}$, all the components of $T_{b}^{a}$ diverge as $\rho \sim
t^{-2p+2p_{3}}$.

For the generalisation of the Kasner metric in the $\mathcal{L}=R^{n}$
theory, (\ref{Xnkasner}), the vacuum terms diverge as $t^{-2n}$. We
therefore require that $2n>P+p_{3}$, or $p_{3}<1$, in order for the vacuum
terms to dominate over the matter terms for a non-comoving perfect fluid,
with general $\gamma $, in the limit $t\rightarrow 0$. For the $\gamma =%
\frac{4}{3},$ comoving perfect fluid ($u_{i}=\delta _{i}^{0}$), the
condition for vacuum domination is $2n>2P-2p_{3}$, or $p_{3}>n-1$. Both of
these conditions are ensured by the boundary values on $p_{3}$ given above.
In the velocity-dominated limit, the vacuum solutions given here are
therefore the appropriate asymptotic solutions on approach to the
singularity, as was found in \cite{Bar06}.

For the generalised Kasner metric in the $\mathcal{L}=(R_{ab}R^{ab})^{n}$
theory, (\ref{Ynkasner}), the vacuum terms diverge as $t^{-4n}$. For vacuum
domination as $t\rightarrow 0$ in this solution we therefore require $%
4n>P+p_{3}$, or $p_{3}<-1+8n-4n^{2}$, for the general non-comoving $\gamma $
fluid; and $4n>2P-2p_{3}$, or $p_{3}>1-6n+4n^{2}$, for the $\gamma =4/3$
comoving fluid. From figure \ref{Yn} it can be seen that solutions of this
class can be in one of two regions, $n_{1}\leqslant n\leqslant n_{2}$ or $%
1/2\leqslant n\leqslant n_{3}$. The validity of the vacuum solution as $%
t\rightarrow 0$ is different in these two different regions. For the region $%
1/2\leqslant n\leqslant n_{3}$ the boundary conditions on the index $p_{3}$
mean that the conditions above are automatically satisfied. For the region $%
n_{1}\leqslant n\leqslant n_{2}$ these inequalities are not always
satisfied. For general $\gamma ,$ the condition for vacuum domination is
only met if $n$ lies in the narrow range $1/6<n<n_{2}$, for any other value
of $n$, in this region, the index $p_{3}$ can be such that the fluid
diverges faster than the vacuum terms. Similarly, for the $\gamma =4/3$
comoving fluid the condition for vacuum domination is only satisfied if $n$
lies in the narrow range $n_{4}<n<n_{2}$, where $n_{4}$ is the real root of
the cubic

\[
4n^{3}-12n^{2}+10n-1=0. 
\]

To understand the evolution of an anisotropic solution of the form (\ref%
{kasner}) in the theory $\mathcal{L}=(R_{ab}R^{ab})^{n}$, where $%
n_{1}\leqslant n\leqslant n_{2}$, it is therefore necessary to take into
account the relativistic motions of any fluid that is present (except in the
narrow ranges of $n$ identified above). This range of $n$ can, however, be
regarded as not belonging to physically interesting theories on other
grounds. In order to agree with weak-field experiments it is necessary for a
gravitational theory to contain a term which approximates the
Einstein-Hilbert action in the weak-field limit (see ref. \cite{Cli05}). Any
theory containing such a limit must therefore have a term in its Lagrangian
that diverges as $t^{-2}$. In considering the behaviour of alternative
theories close to the singularity it is necessary for the extra terms in the
Lagrangian to diverge faster than this if they are to be influential as $%
t\rightarrow 0$. For the theory $\mathcal{L}=(R_{ab}R^{ab})^{n}$ this
requires $n>1/2$. For $n<1/2$ such a term will diverge slower than the
Einstein-Hilbert term which will then dominate and display the standard
Kasner behaviour of general relativity. For the anisotropic solution (\ref%
{Ynkasner}) we are therefore only interested in the region lying in the
range $1/2\leqslant n\leqslant n_{3}$, which has vacuum terms that dominate
the fluid stresses in the velocity-dominated approximation.

It remains to investigate the effects of fluids with stiff equations of
state, $\gamma >4/3$. For sufficiently stiff fluids the velocity-dominated
approximation is not valid and $u^{\alpha }u_{\alpha }\rightarrow 0$ as $%
t\rightarrow 0$ due to the very high inertia of the fluid producing a slow
down under contraction \cite{Bar77b}. In such a limit $u_{0}\rightarrow 1$
and the conservation equations can be solved to give 
\begin{eqnarray*}
&&\rho \sim t^{-\gamma P} \\
&&u_{\alpha }\sim t^{(\gamma -1)P}.
\end{eqnarray*}%
In this approximation we can write $u_{\alpha }u^{\alpha }\sim
(u_{3})^{2}t^{-2p_{3}}\sim t^{2(\gamma -1)P-2p_{3}}$. It can now be seen
that this behaviour occurs when 
\begin{equation}
p_{3_{\ }}\leqslant (\gamma -1)P.  \label{stiff}
\end{equation}%
In this limit it is only required that $\rho $ diverges more slowly than the
vacuum terms, as $\rho u_{\alpha }u^{\alpha }<<\rho $ when $t\rightarrow 0$.

For the solution (\ref{Xnkasner}) to the theory $\mathcal{L}=R^n$ it can be
seen from the condition (\ref{stiff}) that it is necessary for $\gamma
\geqslant 4/3$ in order for the velocity-dominated approximation to break
down (this is derived using the upper limit on $n$ and the lower limit on $%
p_3$). The condition that the fluid stresses diverge more slowly than the
vacuum terms is now 
\begin{equation}
\gamma< \frac{2 n}{ (2 n-1)}
\end{equation}
where it has been assumed $n>1/2$. A sufficient condition to satisfy this
for all allowed values of $n$ is $\gamma < 5/3$. For fluids with a stiffer
equation of state, $5/3 \leqslant \gamma < 2$, the vacuum terms dominate
over the fluid motions provided that $n<1$. For $n<1$, however, the
Einstein-Hilbert term will be the leading one in the gravitational action,
as discussed above.

Similarly, for the solution (\ref{Ynkasner}) to the theory $\mathcal{L}%
=(R_{ab}R^{ab})^{n}$ it can be seen from the condition (\ref{stiff}) that it
is again necessary for $\gamma \geqslant 4/3$ in order for the
velocity-dominated approximation to break down. In the region $1/2\leqslant
n\leqslant n_{3}$ the condition for the vacuum solution to be unperturbed by
the stiff fluid is always satisfied for any fluid with equation of state $%
\gamma <2$. However, in the region $n_{1}<n<n_{2}$ the condition for the
vacuum to diverge faster than the energy density of the stiff fluid is never
satisfied for a fluid which satisfies the necessary condition for the
velocity-dominated approximation to break down (except in the narrow range $%
(7-\sqrt{33})/8<n<n_{2}$ where there are some values of $\gamma $ for which
the vacuum terms can diverge fastest).

\section{Conclusions}

We have investigated some anisotropic cosmological solutions to higher-order
Lagrangian theories of gravity. Whist the standard general relativistic
theory, defined by the Einstein-Hilbert action, appears to be consistent
with all weak-field tests, there is less reason to think that it should be
valid in high curvature regimes, such as in the vicinity of a possible
initial cosmological curvature singularity. In fact, it is in the
high-curvature limit that quantum effects should become important and we
should expect to see deviations from the standard theory. Without knowing
the exact form of such deviations, we have approached this problem by
considering a general class of theories that can be derived from an
arbitrary analytic function the three curvature invariants $R$, $R_{ab}R^{ab}
$ and $R_{abcd}R^{abcd}$. Expanding this function as a power series in these
variables we then expect the dominant term in the Lagrangian to be of the
form $R^{n}$, $(R_{ab}R^{ab})^{n}$ or $(R_{abcd}R^{abcd})^{n}$ as the
singularity is approached. We have found all of the solutions to these
theories that can be expressed in terms of the Bianchi type I line--element (%
\ref{kasner}) in vacuum. These solutions provide simple testing grounds for
the exploration of quantum cosmological effects in higher-order gravity
theory. We have also found the homogeneous, isotropic and spatially flat
solutions to these theories in the presence of a perfect fluid.

Exact vacuum anisotropic solutions of the form (\ref{kasner}) were found for
the theories $\mathcal{L}=R^{n}$ and $\mathcal{L}=(R_{ab}R^{ab})^{n}$,
whilst it was shown that no such solutions exist for theories defined by $%
\mathcal{L}=(R_{abcd}R^{abcd})^{n}$. The properties of these solutions, with
respect to their relation to the more general Bianchi type VIII and IX
cosmological behaviour, has been investigated. We have argued that for all
of the physically relevant new solutions of these theories, the universe
will not experience an infinite number of Mixmaster oscillations as the
singularity is approached. This is an extension of what was found in the
earlier work \cite{Bar06}. The extent to which these vacuum solutions can be
considered as realistic in the presence of a perfect fluid has also been
investigated. It has been shown in the velocity-dominated approximation that
all the anisotropic vacuum solutions found for plausible theories are valid
in the limit $t\rightarrow 0$. The case of stiff fluids that do not satisfy
the velocity-dominated approximation as $t\rightarrow 0$ has also been
investigated. For the solutions to the theories $\mathcal{L}=R^{n}$ it was
found that for a fluid with equation of state $\gamma <5/3$ the vacuum
solutions are good approximations in the vicinity of the singularity. For
the theories $\mathcal{L}=(R_{ab}R^{ab})^{n}$ it was found that the vacuum
solutions are good approximations for all $\gamma <2$.

\ack

T. Clifton acknowledges a PPARC studentship.


\section*{References}

\begin{thebibliography}{99}
\bibitem{Bar83} Barrow J D and Ottewill A C 1983 \textit{J. Phys. A} \textbf{%
16} 2757

\bibitem{schmidt} Schmidt H J 2004 \textit{gr-qc/0407095}

\bibitem{Stelle} Stelle K S 1976 \textit{Phys. Rev. D} \textbf{16} 953

\bibitem{dark1} Carroll S M, De Felice A, Duvvuri V, Easson A, Trodden M and
Turner M S 2005 \textit{Phys. Rev. D} \textbf{71} 063513

\bibitem{dark2} Nojiri S and Odintsov S D 2003 \textit{Phys. Rev. D} \textbf{%
68} 123512

\bibitem{bher} Barrow J D and Hervik S 2006 \textit{Phys. Rev. D} \textbf{73}
023007

\bibitem{bar} Barrow J D 1987 \textit{Phys. Lett. B} \textbf{187} 12

\bibitem{ruz} Ruzmaikin A A and Ruzmaikina T V 1969 \textit{Sov. Phys. JETP} 
\textbf{57} 680

\bibitem{kas} Kasner E 1925 \textit{Trans. Am. Math. Soc.} \textbf{27} 101

\bibitem{bkl} Belinskii V A, Khalatnikov I M and Lifshitz E M 1971 \textit{%
Sov. Phys. Usp.} \textbf{13} 745

\bibitem{misner} Misner C W 1969 \textit{Phys. Rev. Lett.} \textbf{22} 1071

\bibitem{jb1} Barrow J D 1981 \textit{Phys. Rev. Lett.} \textbf{46} 963

\bibitem{jb2} Barrow J D 1982 \textit{Phys. Reports} \textbf{85} 1

\bibitem{chern} Chernoff D and Barrow J D 1983 \textit{Phys. Rev. Lett.} 
\textbf{50} 134

\bibitem{rend} Rendall A D 1997 \textit{Class. Quantum Grav.} \textbf{14}
2341

\bibitem{Bar06} Barrow J D and Clifton T 2006 \textit{Class. Quant. Grav.} 
\textbf{23} L1

\bibitem{dewitt} De Witt B 1965 \textit{The Dynamical Theory of Groups and
Fields} Gordon and Breach, New York

\bibitem{BM} Madsen M and Barrow J D 1989 \textit{Nucl. Phys. B} \textbf{323}
242

\bibitem{Cli05} Clifton T and Barrow J D 2005 \textit{Phys. Rev. D} \textbf{%
72} 123003

\bibitem{schmidt1} Bleyer U and Schmidt H J 1990 \textit{Int. J. Mod. Phys. A%
} \textbf{5} 4671

\bibitem{Car04} Carloni S, Dunsby P K S, Capoziello S and Troisi A 2005 
\textit{Class. Quant. Grav.} \textbf{22} 4839

\bibitem{BK}  Belinskii V A and Khalatnikov I M, 1973 \textit{Sov. Phys.
JETP }\textbf{\ 36} 591

\bibitem{Bar78} Barrow J D 1978 \textit{Nature} \textbf{272} 211

\bibitem{landau} Landau L and Lifshitz E M 1975 \textit{The Classical Theory
of Fields} 4th rev. edn. Pergamon, Oxford

\bibitem{Bar77} Barrow J D 1977 \textit{Mon. Not. R. astron. Soc.} \textbf{%
178} 625

\bibitem{els} Eardley D, Liang E and Sachs R 1972 \textit{J Math Phys} 
\textbf{13} 99

\bibitem{tomita} Tomita K 1975 \textit{Prog Theo Phys} \textbf{54 }730

\bibitem{Bar77b} Barrow J D 1977 \textit{Mon. Not. R. astron. Soc.} \textbf{%
179} 47P
\end{thebibliography}
\end{document}